\newcommand{\be}{\begin{equation}}
\newcommand{\ee}{\end{equation}}
\newcommand{\bea}{\begin{eqnarray}}
\newcommand{\eea}{\end{eqnarray}}
\newcommand{\bean}{\begin{eqnarray*}}
\newcommand{\eean}{\end{eqnarray*}}
\newcommand{\ba}{\begin{array}}
\newcommand{\ea}{\end{array}}

\newcommand{\bc}{\begin{center}}
\newcommand{\ec}{\end{center}}
\newcommand{\bi}{\begin{itemize}}
\newcommand{\ei}{\end{itemize}}

\newcommand{\ra}{\rightarrow}

\newcommand{\aqt}{{\cal A}_{TT}(Q_T)}
\newcommand{\aqtn}{{\cal A}_{TT}}

\documentclass{PoS}

\title{Soft gluon resummation in Drell-Yan dilepton 
production at small transverse momentum: \\
spin asymmetry and a novel asymptotic formula}

\ShortTitle{Soft gluon resummation in 
Drell-Yan dilepton production}

\author{\speaker{Hiroyuki Kawamura}\\
        Radiation Laboratory, RIKEN,
        Wako 351-0198, Japan\\
        E-mail: \email{hiroyuki@ribf.riken.jp}}

\author{Jiro Kodaira\thanks{Deceased.}\\
        Theory Division, KEK, Tsukuba 305-0801, Japan}

\author{Kazuhiro Tanaka\\
        Department of Physics, Juntendo University,
        Inba, Chiba 270-1695, Japan\\
        E-mail:\email{tanakak@sakura.juntendo.ac.jp}}

	\abstract{We discuss the double-spin asymmetry $\aqt$ in the transversely 
          polarized Drell-Yan process at small transverse-momentum $Q_T$
          of the produced dilepton.
          The soft gluon corrections relevant for small $Q_T$ are resummed to 
          all orders in $\alpha_s$, up to the next-to-leading logarithmic accuracy. 
          We show that the soft gluon corrections largely cancel in the spin
          asymmetry, but the significant corrections still remain.
          The asymmetries $\aqt$ are calculated for $pp$ collision at
          RHIC and J-PARC, and for $p\bar{p}$ collision at GSI.
          A novel asymptotic formula for $\aqt$ at small $Q_T$ is presented, 
          which provides a new approach to extract the transversity $\delta q(x)$ 
          from the experimental data.}

\FullConference{8th International Symposium on Radiative Corrections (RADCOR)\\
		 October 1-5 2007\\
		 Florence, Italy}

\begin{document} 

\section{Introduction} 
The transversely polarized Drell-Yan process (tDY),
$p^\uparrow + p^\uparrow\ra l+\bar{l}+X$, 
$p^\uparrow + {\bar{p}^\uparrow} \ra l+\bar{l}+X$, 
is one of the processes where we can access the chiral-odd 
transversity distributions $\delta q(x)$. 
It can be measured at the ongoing experiments at RHIC and 
the possible future experiments at J-PARC, GSI, etc. 
In this article, we consider the double-spin asymmetry for tDY,
especially at the small transverse momentum $Q_T$ of the dilepton, 
where the bulk of the production occurs. 
When $Q_T$ is much smaller than the invariant mass of the dilepton, $Q$, 
the soft gluon corrections proportional to   
$\alpha_s^n\ln^m(Q^2/Q_T^2)$ ($m\leq 2n-1$) are enhanced and 
have to be resummed to all orders of perturbation theory 
to obtain a reliable prediction.
In \cite{KKT:07,KKT:08}, 
we studied the double transverse-spin asymmetry $\aqt$ at small $Q_T$  
using the result of \cite{KKST:06}, 
where the $Q_T$ spectrum of the dilepton for tDY was calculated
performing the soft gluon resummation up to the next-to-leading 
logarithmic (NLL) accuracy.
In the following, we will summarize the main results obtained in those works.
 
\section{The double-spin asymmetry for tDY at a measured $Q_T$}
The spin-dependent 
($\Delta_T d \sigma \equiv (d\sigma^{\uparrow\uparrow}-d\sigma^{\uparrow\downarrow})/2$)
and spin-independent 
($d \sigma \equiv (d\sigma^{\uparrow\uparrow}+d\sigma^{\uparrow\downarrow})/2$)
parts of
the tDY cross section with the NLL soft gluon resummation are expressed as 
\be
\frac{\left( \Delta_T\right)d\sigma}{dQ^2dQ_T^2dyd\phi} =
\left( \cos( 2\phi)/2 \right)
\frac{2 \alpha^2}{3\, N_c\, S\, Q^2}
\biggl[ \left(\Delta_T\right) \tilde{X}^{\rm NLL} (Q_T^2 , Q^2,y)
+ \left( \Delta_T\right) \tilde{Y} (Q_T^2 , Q^2,y)\biggr] \ ,
\label{NLL+LO}
\ee
where $\sqrt{S}$ and $y$ are the total energy and dilepton's rapidity 
in the hadron CM system, and the factor $\cos(2\phi)/2$ for  
the spin-dependent part shows the characteristic dependence on  
the azimuthal angle $\phi$ of one of the outgoing leptons 
with respect to the incoming nucleon's spin axis.
The first term ($\Delta_T\tilde{X}^{\rm NLL}$ and $\tilde{X}^{\rm NLL}$) 
corresponds to the NLL resummed component 
which contains the first three towers of the logarithmically-enhanced corrections, 
$\alpha_s^n\ln^m(Q^2/Q_T^2)/Q_T^2~(m=2n-1,2n-2,$ and  $2n-3)$, 
and the second term ($\Delta_T\tilde{Y}$ and $\tilde{Y}$)  
is the remaining non-enhanced component 
of $O(\alpha_s)$ and is determined 
such that the expansion of (\ref{NLL+LO}) to $O(\alpha_s)$ reproduces 
the corresponding leading-order (LO) cross section for $Q_T>0$. 
Accordingly, we call (\ref{NLL+LO}) the ``NLL+LO'' cross sections.  
$\Delta_T \tilde{X}^{\rm NLL}$ and $\tilde{X}^{\rm NLL}$ 
are obtained through various elaborations
\cite{BCDeG:03,KKST:06, KKT:07} of the Collins-Soper-Sterman 
resummation formalism \cite{CSS:85}, 
and $\Delta_T\tilde{X}^{\rm NLL}$ is given by the integral over 
the impact parameter $b$, conjugate to $Q_T$, as
\bea
\Delta_T \tilde{X}^{\rm NLL} (Q_T^2, Q^2, y)
&&\!\!\!\!\!\!\!=
\int_{\cal C} db\frac{b}{2} J_0(b Q_T)e^{S (b,Q)-g_{NP}b^2} 
\left[~\delta H \left(x^0_1,x^0_2;\ \frac{b_0^2}{b^2}\right)\right.
\nonumber\\
&&\hspace*{0.5cm}+ 
\left. \frac{\alpha_s (Q^2)}{2\pi} \left\{
\int_{x^0_1}^1  \frac{dz}{z}  \Delta_T C_{qq}^{(1)}(z)
\delta H \left(\frac{x_1^0}{z},x_2^0;\ \frac{b_0^2}{b^2}\right)
+ \left(x_1^0 \leftrightarrow x_2^0 \right)
\right\} \right],
\label{resum}
\eea 
where the DY scaling variables are denoted as $x_{1,2}^0=\sqrt{Q^2/S}e^{\pm y}$, 
$J_0(bQ_T)$ is a Bessel function and $b_0=2e^{-\gamma_E}$ with
$\gamma_E$ the Euler constant.
$\delta H$ denotes the sum of the products of the NLO transversity
distributions: 
$\delta H(x_1,x_2; \mu^2)
=\sum_qe_q^2\left[\delta q(x_1,\mu^2)\delta \bar{q}(x_2,\mu^2)+
(x_1\leftrightarrow x_2)\right]$ for $pp$ collision
and 
$\delta H(x_1,x_2; \mu^2)
=\sum_qe_q^2\left[\delta q(x_1,\mu^2)\delta q(x_2,\mu^2)+
\delta \bar{q}(x_1,\mu^2)\delta \bar{q}(x_2,\mu^2)\right]$
for $p\bar{p}$ collision.
Note that there is no gluon transversity distribution.
The Sudakov factor $e^{S(b,Q)}$ resums the enhanced logarithmic
corrections up to the NLL level and is common for the polarized and 
unpolarized cross sections.
Using $\lambda=\beta_0\alpha_s(Q^2)\ln(Q^2b^2/b_0^2+1)$ 
with $\beta_0=(11N_c-2N_f)/(12\pi)$,
the exponent is obtained as 
$S(b,Q)=\frac{1}{\alpha_s (Q^2) }h^{(0)}(\lambda)+h^{(1)}(\lambda)$, 
where
$h^{(0)}(\lambda)=(A_q^{(1)}/2\pi\beta_0^2)[\lambda+\ln(1-\lambda)]$ 
with $A_{q}^{(1)}=2C_F=(N_c^2-1)/N_c$ 
collects the LL contributions and $h^{(1)}(\lambda)$
collects the NLL contributions. The explicit form of $h^{(1)}(\lambda)$ 
as well as the coefficient function at the NLL level,
$\Delta_TC^{(1)}_{qq}(z)$, is found in \cite{KKST:06}.
The $b$ dependence of $\delta H(x_1,x_2 ; b_0^2/b^2)$ 
associated with the NLO evolution of the parton distributions 
is also organized in terms of $\lambda$ \cite{BCDeG:03, KKT:07}.
The contour ${\cal C}$ in (\ref{resum}) is taken in the complex $b$ space
to avoid the singularity at $\lambda=1$ (see \cite{KKST:06,KKT:07}) and 
we have introduced a Gaussian smearing function $e^{-g_{NP}b^2}$ 
with a parameter $g_{NP}$ to complement the nonperturbative effects 
from the extremely large $|b|$ region.   
The resummed component of the unpolarized cross section, 
$\tilde{X}^{\rm NLL}$, is obtained similarly. 
The explicit expressions for 
$\tilde{X}^{\rm NLL}$, $\Delta_T\tilde{Y}$ and $\tilde{Y}$ are 
found in \cite{KKT:07,KKST:06}. 

\begin{figure}
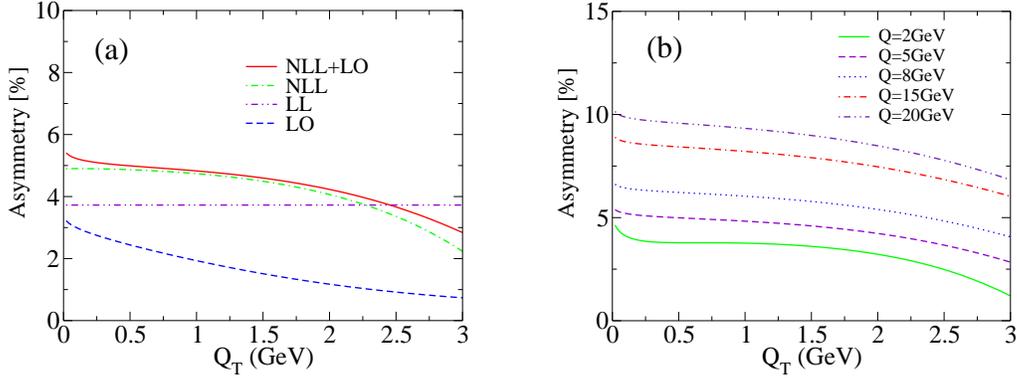

\bc
\includegraphics[height=5cm]{RHIC_200_5_y2_asym_2.eps}~~~~~~~~~~~
\includegraphics[height=5cm]{RHIC_200_y2_asym.eps}
\ec
\caption{The asymmetries at $\phi=0$ for $pp$ collision with 
RHIC kinematics, $\sqrt{S}=200$ GeV and $y=2$, in (a) various accuracy
for $Q=5$ GeV and (b) the NLL+LO accuracy for various $Q$.}
\label{fig:1}
\end{figure}

The NLL+LO asymmetry $\aqt$ is defined by 
\bea
\hspace*{-0.5cm}
\aqt=\frac{\Delta_Td\sigma/dQ^2dQ_T^2dyd\phi}{d\sigma/dQ^2dQ_T^2dyd\phi}
=\frac{1}{2} \cos(2\phi)
\frac{\Delta_T\tilde{X}^{\rm NLL}(Q_T^2 , Q^2, y)
+\Delta_T\tilde{Y}(Q_T^2, Q^2, y)}
{\tilde{X}^{\rm NLL}(Q_T^2, Q^2, y)+\tilde{Y}(Q_T^2, Q^2, y)}\ ,
\label{asym}
\eea
for measured $Q_T$, $Q$, $y$ and $\phi$. 
When the resummed components are expanded to $O(\alpha_s)$, 
the above asymmetry reduces to the LO prediction $\aqtn^{\rm LO}(Q_T)$ for $Q_T>0$.  
For comparison, we also introduce the NLL asymmetry 
$\aqtn^{\rm NLL}(Q_T)=[\cos(2\phi)/2]
\Delta_T\tilde{X}^{\rm NLL}/\tilde{X}^{\rm NLL}$
and the LL asymmetry
$\aqtn^{\rm LL}=[\cos(2\phi)/2]\delta H(x_1^0,x_2^0;Q^2)/H(x_1^0,x_2^0;Q^2)$,
where the latter is obtained from $\aqtn^{\rm NLL}(Q_T)$ by dropping 
the NLL terms and, for $pp$ collision, 
$H(x_1,x_2; \mu^2)=\sum_qe_q^2\left[q(x_1,\mu^2)\bar{q}(x_2,\mu^2)
+(x_1\leftrightarrow x_2)\right]$.
Note that $\aqtn^{\rm LL}$ does not depend on $Q_T$ since the soft
gluon corrections at the LL level cancel out in the asymmetry \cite{KKT:07}.

In Fig.~1(a), we compare these asymmetries in various accuracy 
for $pp$ collision at RHIC with $\sqrt{S}=200$GeV, $Q=5$GeV, $y=2$. 
Here and below, we use the NLO transversity constructed as in \cite{MSSV:98} 
and $g_{NP}=0.5$GeV$^2$, and set $\phi=0$.
The NLL+LO asymmetry $\aqt$ is flat in the small $Q_T$ region 
and is close to $\aqtn^{\rm NLL}(Q_T)$; i.e., $\aqt$ is dominated by the resummed
components, $\Delta_T\tilde{X}^{\rm NLL}$ and $\tilde{X}^{\rm NLL}$. 
The flat behavior of $\aqtn^{\rm NLL}(Q_T)$ is due to the universality
of the soft gluon corrections in the Sudakov factor $e^{S(b,Q)}$ up to the NLL level. 
Compared with $\aqtn^{\rm LL}$, $\aqtn^{\rm NLL}(Q_T)$ is enhanced 
by the contributions at the NLL level.
The LO asymmetry $\aqtn^{\rm LO}(Q_T)$ is much smaller than the other asymmetries, 
indicating that the soft gluon resummation is crucial for the prediction 
of the asymmetry. 
Moreover, the NLL+LO asymmetry $\aqt$ is larger than the asymmetry for 
the $Q_T$-integrated cross sections 
\begin{equation}
A_{TT}\equiv\frac{\int dQ_T^2 \left( \Delta_T d\sigma /dQ^2dQ_T^2dyd\phi \right)}
{\int dQ_T^2 \left( d\sigma /dQ^2dQ_T^2dyd\phi \right)}
=\frac{1}{2}\cos(2 \phi) \frac{\delta H(x_1^0, x_2^0 ; Q^2)+\cdots}
{H(x_1^0, x_2^0 ; Q^2)+\cdots}\ .
\label{eq:att}
\end{equation}
Indeed, $A_{TT}= 4.0\%$ in the present case.  
In the RHS of (\ref{eq:att}), the ellipses represent the NLO correction
terms, and $A_{TT}$ coincides with the NLO asymmetry calculated in \cite{MSSV:98},
because our differential cross sections (\ref{NLL+LO}) are constructed 
to satisfy the ``unitarity constraint'' \cite{BCDeG:03}. 
This also implies $A_{TT}\simeq\aqtn^{\rm LL}$.
The NLL+LO asymmetries $\aqt$ for different values of $Q$ are shown in Fig.1(b).
$\aqt$ is flat in all cases and increases as $Q$ increases. 
This $Q$ dependence is a result of the small-$x$ behavior of the relevant 
parton distributions, in particular, the steep rise of the unpolarized 
sea distributions at small $x_{1,2}^0=\sqrt{Q/S}e^{\pm y}$, which
enhances the denominator of (\ref{asym}) for small $Q$.

\begin{figure}
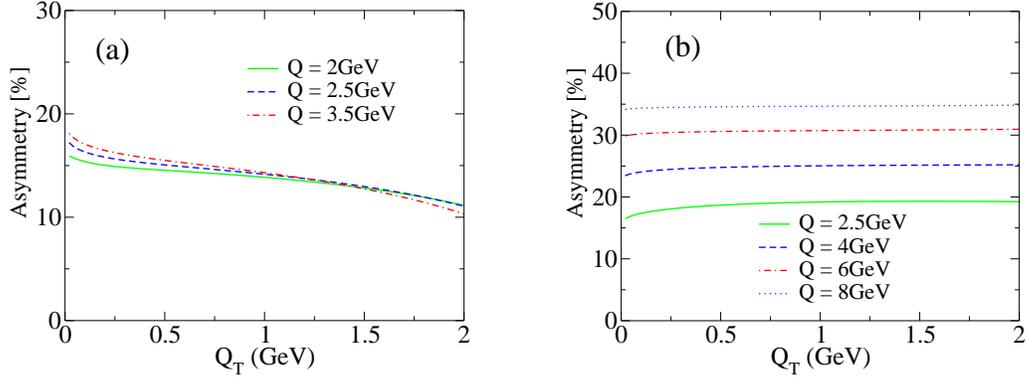

\bc
\includegraphics[height=5cm]{J-PARC_10_y0_asym.eps}~~~~~~~~~~~~
\includegraphics[height=5cm]{GSI_14.5_y0_asym.eps}
\ec
\caption{The NLL+LO $\aqt$  of (2.3) at $\phi=0$ for (a) $pp$ collision 
with J-PARC kinematics, $\sqrt{S}=10$ GeV and $y=0$, and 
(b) $p\bar{p}$ collision with GSI kinematics, $\sqrt{S}=14.5$ GeV 
and $y=0$.}
\label{fig:2}
\end{figure}

Figure 2(a) shows the NLL+LO asymmetries $\aqt$ for $pp$ collision at J-PARC
with $\sqrt{S}=10$GeV, $y=0$. $\aqt$ are about 15\%,
irrespective of the value of $Q$, and are larger than those for the RHIC case, 
since the distributions at the moderate $x$ values are probed at J-PARC
\cite{KKT:07}. In all the cases in Figs.1 and 2(a) for $pp$ collision, 
$\aqt$ are larger by 15-30\% than $A_{TT}\simeq\aqtn^{\rm LL}$. 
Figure 2(b) shows the NLL+LO $\aqt$ for $p\bar{p}$ collision at GSI 
with $\sqrt{S}=14.5$ GeV, $y=0$. 
The asymmetries are much larger than the previous two cases,
since the products of {\em valence} distributions are probed at {\em moderate} $x$. 
This fact also leads to $\aqt\simeq A_{TT}\simeq\aqtn^{\rm LL}$ at GSI 
(see \cite{KKT:08} for the detail).

\section{The saddle-point formula for tDY asymmetries at NLL}
For all cases discussed in Figs.1(b) and (2), we actually have 
$\aqt\simeq\aqtn^{\rm NLL}(Q_T)\simeq\aqtn^{\rm NLL}(Q_T=0)$, 
similarly as in Fig.1(a).
At $Q_T=0$, the $b$ integral in (\ref{resum}) 
is controlled by a saddle-point and can be evaluated analytically 
\cite{KKT:07} as
\be
\Delta_T\tilde{X}^{\rm NLL} (0 , Q^2, y)
=\left[\frac{b_0^2}{4Q^2 \beta_0 \alpha_s(Q^2)} 
\sqrt{\frac{2\pi}{{\zeta^{(0)}}''(\lambda_{SP})}} 
e^{-\zeta^{(0)}(\lambda_{SP})+h^{(1)}(\lambda_{SP})} \right] 
\delta H \left(x_1^0,x_2^0; \frac{b_0^2}{b_{SP}^2}\right) ,
\label{eq:speval}
\ee 
where 
$\zeta^{(0)}(\lambda) \equiv - \lambda/[\beta_0 \alpha_s(Q^2)]
-h^{(0)}(\lambda)/ \alpha_s (Q^2)
+ [g_{NP}b_0^2 /Q^2 ]e^{\lambda/[ \beta_0 \alpha_s (Q^2 )]}$,
and the saddle-point value, 
$\lambda_{SP}=\beta_0\alpha_s( Q^2 ) \ln(Q^2b_{SP}^2/b_0^2+1)
\simeq \beta_0\alpha_s( Q^2 ) \ln(Q^2b_{SP}^2/b_0^2 )$,
is determined by the condition ${\zeta^{(0)}}' (\lambda_{SP} )=0$.
Note that the saddle-point formula (\ref{eq:speval}) is exact up to 
$O(\alpha_s)$ corrections that actually correspond to the NNLL
contributions in the $Q_T\approx 0$ region \cite{KKT:07}. 
Similarly, the saddle-point formula for $\tilde{X}^{\rm NLL}$ is 
given by (\ref{eq:speval}) with $\delta H(x_1^0, x_2^0;\ b_0^2/b_{SP}^2)$ 
replaced by $H(x_1^0, x_2^0;\ b_0^2/b_{SP}^2)$.
Therefore, the large radiative corrections included in the square bracket 
in the RHS of (\ref{eq:speval}) cancel out in the asymmetry, and 
we obtain a remarkably compact formula using the $Q_T\ra 0$ limit 
of the NLL asymmetry:
\begin{equation}
\aqt\simeq\aqtn^{\rm NLL}(Q_T =0)=\frac{1}{2}\cos(2 \phi ) 
\frac{\delta H \left( x_1^0, x_2^0;\ b_0^2 /b_{SP}^2 \right)}
{H \left( x_1^0, x_2^0;\ b_0^2 /b_{SP}^2 \right)}\ .
\label{eq:attnll}
\end{equation}
In the RHS, the corrections at the NLL level which survive the cancellation 
are entirely absorbed into the unconventional scale $b_0/b_{SP}$ in the relevant 
distribution functions, 
and $b_0/b_{SP}\simeq 1$GeV for all cases in Figs.1 and 2 \cite{KKT:07,KKT:08}.
This scale $b_0/b_{SP}$ in (\ref{eq:attnll}), instead of $Q$, 
explains why $\aqt$ is always larger than $\aqtn^{\rm LL}$
or $A_{TT}$ of (\ref{eq:att}) for $pp$ collision \cite{KKT:07}.
Also, this formula clarifies that the $Q$ dependence of $\aqt$ 
in Figs.1(b) and 2 reflects the shape of the parton distributions 
at the scale $b_0/b_{SP}\simeq 1$ GeV. 
The saddle-point formula (\ref{eq:attnll}) approaches 
the exact asymmetry in the $Q\rightarrow\infty$ limit.

In Table 1, 
we list the values of $\aqtn^{\rm NLL}(Q_T=0)$ obtained from 
the saddle-point formula (\ref{eq:attnll}) for the kinematics in Figs.1 and 2. 
The formula indeed reproduces the NLL+LO $\aqt$ in the flat region to 
the 10\% accuracy, i.e., to the canonical size of the ${\cal O}(\alpha_s)$ 
corrections associated with the NLL accuracy. 
It has been demonstrated that certain NNLL corrections to (\ref{eq:attnll}) can 
grow for the RHIC kinematics at small $Q$, corresponding to the small-$x$ region, 
but those corrections are always small for J-PARC and GSI \cite{KKT:07,KKT:08}.
Therefore, the formula (\ref{eq:attnll}) is particularly useful for 
J-PARC and GSI to extract the transversity distributions directly from the data.

The present framework for the soft gluon resummation can be applied to
other polarized and unpolarized processes,
such as vector boson production at RHIC, LHC, etc. 
   
\begin{table}
\centerline{
\begin{tabular}{|c|c|c|c||c|c|c||c|c|c|}
\hline
& \multicolumn{3}{|c||}{RHIC {\small ($\sqrt{S}=200$GeV,$y=2$)}}& 
\multicolumn{3}{|c||}{J-PARC {\small ($\sqrt{S}=10$GeV,$y=0$})}& 
\multicolumn{3}{|c|}{GSI {\small($\sqrt{S}=14.5$GeV,$y=0$)}}\\
\hline
$Q$ 
& 5GeV  & 8GeV & 15GeV  
& 2GeV & 2.5GeV  & 3.5GeV & 2.5GeV & 4GeV& 6GeV\\
\hline
Eq.(3.2)
& 5.4\% & 6.6\% & 8.7\% 
& 14.1\% & 14.5\% & 14.8\%  & 20.2\% & 25.6\% & 30.9\% \\
\hline
 \end{tabular}}
\caption{$\aqtn^{\rm NLL}(Q_T =0)$ at $\phi=0$ using the saddle-point
 formula (3.2).}
\label{tab.1}
\end{table}

\bigskip
This work is supported by the Grant-in-Aid 
for Scientific Research No.~B-19340063.


\end{document}